\documentclass{article}
\PassOptionsToPackage{numbers, compress}{natbib}

\usepackage[preprint]{neurips_data_2021}
\usepackage{makecell, cellspace, caption}

\usepackage{wrapfig,lipsum,booktabs}
\usepackage{amsmath,amssymb,amsfonts}
\usepackage{algorithmic}
\usepackage{graphicx}
\usepackage{textcomp}
\usepackage{wrapfig}
\usepackage{tcolorbox}

\usepackage{hyperref}
\hypersetup{
    colorlinks=true,
    linkcolor=blue,
    filecolor=magenta,      
    urlcolor=black,
    citecolor=black
}
\usepackage{lipsum}
\def\BibTeX{{\rm B\kern-.05em{\sc i\kern-.025em b}\kern-.08emT\kern-.1667em\lower.7ex\hbox{E}\kern-.125emX}}
\definecolor{Gray}{gray}{0.3}

\usepackage[utf8]{inputenc} 
\usepackage[T1]{fontenc}    
\usepackage{hyperref}       
\usepackage{url}            
\usepackage{booktabs}       
\usepackage{amsfonts}       
\usepackage{nicefrac}       
\usepackage{microtype}      
\usepackage{xcolor}         

\title{AugmentedCode: Examining the Effects of Natural Language Resources in Code Retrieval Models}

\author{Mehdi Bahrami$^{1}$
\And
N.C. Shrikanth$^{2}$
\And
Yuji Mizobuchi$^{3}$
\And
Lei Liu$^{1}$
\And
Masahiro Fukuyori$^{3}$
\And
Wei-Peng Chen$^{1}$
\And
Kazuki Munakata$^{3}$
\And
\\ \\
$^{1}$Fujitsu Research of America, Sunnyvale, CA, USA
\\
$^{2}$North Carolina State University, Raleigh, NC, USA
\\
$^{3}$Fujitsu Research Ltd., Kawasaki, Japan
\\ \\
\texttt{\{mbahrami,mizobuchi.yuji\}@fujitsu.com}
\\
\texttt{snaraya7@ncsu.edu}
\\
\texttt{\{lliu,wchen,fukuyori,munakata\.kazuki\}@fujitsu.com}
\\
}
\begin{document}
\maketitle
\begin{abstract}
Code retrieval is allowing software engineers to search codes through a natural language query, which relies on both  natural language processing and software engineering techniques.
There have been several attempts on code retrieval from searching snippet codes to function codes. In this paper, we introduce Augmented Code (AugmentedCode) retrieval which takes advantage of existing information within the code and constructs augmented programming language to improve the code retrieval models' performance. We curated a large corpus of Python and showcased the the framework and the results of augmented programming language which outperforms on CodeSearchNet and CodeBERT with a Mean Reciprocal Rank (MRR) of 0.73 and 0.96, respectively. The outperformed fine-tuned augmented code retrieval model is published in HuggingFace at \href{https://huggingface.co/Fujitsu/AugCode}{https://huggingface.co/Fujitsu/AugCode} and a demonstration video is available at: \href{https://youtu.be/mnZrUTANjGs}{https://youtu.be/mnZrUTANjGs}.
\end{abstract}


\section{Introduction}
A frequent practice for software developers (especially novices) is to find specific source code knowledge internally (within the organization) or externally\footnote{https://stackoverflow.com/\\\footnotesize{\par* Equal contribution}}~\cite{singer2010examination,sadowski2015developers}. Recently (2020) Liu et al. report an emerging trend in both research and code retrieval tools~\cite{liu2020opportunities}. Although the current target users of this technology are software engineers, we foresee the target users could be largely replaced by machines (bots) who will be able to synthesize applications largely using code retrieval models~\cite{lebeuf2017software}. Mainly search engines have been used for retrieving codes with some limitation such as lack of retrieving a function code based imported libraries. 
In this paper, to address these issues, we introduce augmented code approach that look into the code and related resources from different perspectives to improve machine-learning based code retrieval.  
The contributions of this work include: i) construction of different augmented codes based on different code attributes; ii) analyse and evaluate the augmented codes based on the state-of-the-art machine-learning based code search architectures.
\subsection{Related work}
Recent studies on natural language processing show that unsupervised pre-trained models, created from a large corpus, improve accuracies across a variety of downstream tasks. For example, ELMo\cite{peters2018deep}, GPT\cite{radford2018improving}, and BERT\cite{devlin2018bert} have achieved significant improvements over the existing approaches. 

In terms of the programming domain, CodeBERT\cite{feng2020codebert}, which is based on Roberta\cite{liu2019roberta} architecture with a corpus of programming languages, has outperformed over CodeSearchNet architecture for a code retrieval task. The model is pre-trained with the CodeSearchNet\cite{husain2019codesearchnet} corpus and is fine-tuned as a sentence pair classification task, where one of the sentences is a code and the other is the description of the code.
%
Cambronero et al.\cite{cambronero2019deep} embed the query and the codes into a shared space by applying several deep learning architectures and rank the codes in a order of distance. 
However, none of the previous studies focus on different attributes of code search and their effects on the performance of the trained models. In this paper we use CodeBERT and CodeSearchNet to evaluate different attributes of codes.
\section{Augmented Code Framework}
\label{sec:augmented code framework}
\begin{figure*}[h]
    \centering
    \includegraphics[width=15cm]{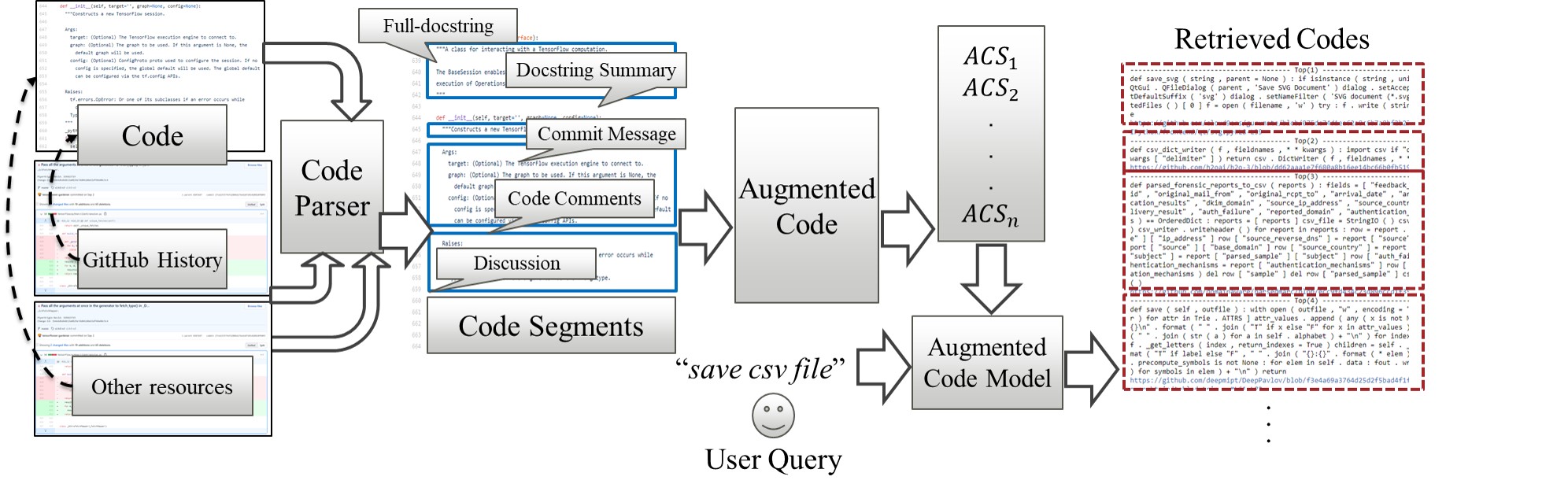}
    \caption{Overview of Augmented Code Framework}
    \label{fig:overview}
\end{figure*}

Figure~\ref{fig:overview} shows the overview of Augmented Code framework. The original source-code can be decomposed into different segments (some of them listed in Figure~\ref{fig:overview}) that include class name, function name, full-code's description (docstring), a summary of code's description, parameters, arguments, line of codes, code comments. In addition, we also take into account of data curation from other resources such as GitHub~\footnote{https://github.com/} history that includes users' commit messages and forum discussion on a particular code. We use a heuristic approach to construct different Augmented-Code Scenarios ($ACS$) where each scenario constructed from different code segments. Each $ACS$ can be used to train different machine-learning models where the trained model aims to retrieve a code per user's natural language query. 
\subsection{Why}
\subsubsection{Objective} To retrieve (predict) an appropriate source-code (method or function) requested per user's natural language query where the target user is a software engineer and target source-code is a large corpus of any programming language (Python in this study).  

\subsubsection{Goal} Productivity (production-rate) in Software Engineering is the number of lines of code over time~\cite{endres2003handbook}. Thus better code retrieval models can aid developers in practice to retrieve useful code more frequently to satisfy to their requirements. In other words, more lines of code being pushed more often to production. That could cascade to minimize release duration and enable faster time to market.

At ICSE 2012, Hindle et al. conjectured that 

``\textit{..most software is also natural, in the sense that it is created by humans at work, with all the attendant constraints and limitations - and thus, like natural language, it is also likely to be repetitive and predictable}''~\cite{hindle2012naturalness}

Thus in those lines, this research investigates a related conjecture that there exist certain natural language assets which are associated with that code which is often under-explored in the code retrieval context.

Hence, we ask:

\begin{tcolorbox}[colback=gray!5,colframe=gray!50]
\textbf{Can we improve code retrieval models, with the underlying natural language resources?}  
\end{tcolorbox} 

By natural language resources we mean, \textit{code comments}, \textit{docstrings} and \textit{commit messages} that have an association with the code. To answer our central question we run numerous experiments with different scenarios that utilize different parts of the associate natural language assets. For instance, if the results of our experiments turns out to value \textit{code comments} more than \textit{docstrings} or \textit{commit-messages}, then in the absence of \textit{code comments} researchers may find ways to augment poorly documented code with synthesized comments. Such a reasoning would help researchers to focus on a specific direction (such as code comment syntheses) to build better code search models. Despite a fairly recent area, numerous studies such as ~\cite{hu2018deep,sridhara2011automatically} have shown pragmatic approaches to synthesize documentation for high-level instructions.

Additionally, the methods discussed in this article portray the simplicity to augment code retrieval based models with underlying natural language data like \textit{code comments}. In other words, the methods and associated assets of this work can serve as a guide to practitioners. If they wish to either augment their existing code search models or build a code search model from scratch.

\subsection{What}
\subsubsection{Docstring}
In python programming, ``\textit{A docstring is a string literal that occurs as the first statement in a module, function, class, or method definition. Such a docstring becomes the \_\_doc\_\_ special attribute of that object.}''~\footnote{https://www.python.org/dev/peps/pep-0257/\#what-is-a-docstring}. A docstring has a high-level short description about the code followed by some examples (optional) with return type and exception/error information appended. Additionally, docstrings can be written in various styles (formats) like Google, NumPy/SciPy, reStructured Text, and Epytext.

\subsubsection{Code comments} An annotation added by the developer in a human-readable language which is part of the source code but ignored by the compiler. Unlike docstrings, code comments are not structured (does not adhere to a template) and usually describe a single objective, typically covering few lines of code.

\subsubsection{Commit Message} A log message describing the changes to push into a version-control system. The commit messages are written typically in natural language by the owner (software developer). By default popular distributed version control system like GitHub~\footnote{https://www.git-scm.com/docs/git-commit} does not restrict developers to write in a specific format.  Having said like \textit{code comments}, well written and concise `Commit messages' can ease code-review practices. For instance, defect prediction models parse commit messages to trace defect inducing changes~\cite{shrikanth2020assessing}. We retrieve commit message using GitHub API~\footnote{https://docs.github.com/en/free-pro-team@latest/rest} using repository and commit hash information available in the CodeSearchNet default data. 

\begin{table*}[h!]
\small
\centering
\caption{5 Augmented-Code Scenarios (ACS) to build code retrieval models ($X  \Longrightarrow Y$) in CodeSearchNet and CodeBERT architectures (ACS=0 is the default case for CodeSearchNet and CodeBERT, whereas our results endorse ACS=4).  }
\begin{tabular}{|l|l|l|}
\hline
\label{tbl:scenarios}
\textbf{ACS} & \textbf{X (attribute:$docstring\_tokens$)}                   & \textbf{Y (attribute:$code\_tokens$)}                           \\ \hline

\textbf{0 (default)} & Tokenized short description of docstring  & Tokenized (code and code comments) \\ \hline
1                 & Tokenized code comments             &     Tokenized code                       \\

2                 & Tokenized (code comments  and entire docstring)                                                & Tokenized code                                                   \\

3                 & Tokenized (code comments, entire docstring & Tokenized (code and  code comments)\\ 
& and commit message) & \\  

4                 & Tokenized (code comments  and entire docstring)    & Tokenized (code and code comments)                    \\

5                 &  Tokenized short description of docstring                 & Tokenized code                             \\ \hline
\end{tabular}
\end{table*}

\subsection{How}
To understand the importance of different natural language entities like \textit{code comments}, \textit{docstrings} and \textit{commit messages} we build and assess various machine learning models on recent architectures specifically \textit{CodeSearchNet} and \textit{CodeBERT}. This allows us to compare and debate directly our models with state of the art techniques. 

Later in Section~\ref{sec:experiment} we assess 17 machine learning models listed in Table~\ref{tbl:experiment_results} based on five different augmented code scenarios (ACS). The five scenarios listed in Table~\ref{tbl:scenarios} use combinations \textit{code comments}, \textit{docstrings} and \textit{commit messages} to investigate our central question. We also previously published PyTorrent~\cite{bahrami2021pytorrent} which includes 218,814 Python software package with more than 655M Line of Codes (LoC). PyTorrent is made available public here~\cite{pytorrent-zenodo},\cite{pytorrent_github} and a Python language model ~\cite{pytorrent-v1-model}\footnote{PyTorrent Dataset: \url{https://zenodo.org/record/4546290}}.

As mentioned earlier, we heavily reuse state of the art architectures like CodeSearchNet and CodeBERT to build various code retrieval models. To achieve that we transform our data to suit CodeSearchNet training data schema (available here~\footnote{https://github.com/github/CodeSearchNet}). That schema currently lists 12 attributes of which \textit{docstring\_tokens}  (`X') and \textit{code\_tokens}  ('Y') are required attributes. To `X' and `Y' we append natural language resources such as \textit{code comments}, \textit{commit messages} and \textit{docstrings} to create 5 + 1 (default) augmented code scenarios (\textit{$ACS$}) which is listed in Table~\ref{tbl:scenarios}.
Since CodeSearchNet is designed to build models for function, docstring pairs mined from Github repositories attributes in that training schema specifically; repo, path, and URL are set to python package name, python script path and not applicable.





\section{Experiment}\label{sec:experiment}

Before we can show that our proposed model is scalable, we have to show that we can both reproduce and augment code retrieval models using the state of the art CodeSearchNet challenge \cite{husain2019codesearchnet} and CodeBERT \cite{feng2020codebert}. Hence, we perform two experiments. First, we evaluate using Mean Reciprocal Rank (MRR) on each of the 999 distractor snippets on default and augmented models. Lastly, we evaluate the best performing (highest MRR) augmented model on a larger dataset that we curated to portray its robustness.


\subsection{Results}
Table~\ref{tbl:experiment_results} shows the experimental results of different augmented programming language codes on two different architectures of CodeSearchNet (CSNet) and CodeBERT. The detail of different architectures i.e., Self-attention - SelfAtt) explained in \cite{feng2020codebert} and \cite{husain2019codesearchnet}. \textit{$ACS$=0} refer to augmented scenario and in this case is scenario \#0 which is the default data which has been used in both CodeSearchNet and CodeBERT. As another example, \textit{$ACS$=4} refer to an augmented programming language scenario where it is augmented developer code comments and code raw code comments. As shown in this experiment, the performance of code retrieval is improved on both architectures when we applied the augmented programming languages (e.g.  \textit{$ACS$=4}).\\
Although the previous studies have ignored these additional resources, our results of different augmented codes show that adding any additional code attributes to a code search model, it improves the performance of each model significantly.
\begin{table}[ht]
    \centering
    \caption{Code Retrieval performance (measured in Mean Reciprocal Rank) of different augmented programming language codes. (ACS=0 is the default case for replication and  ACS=4 yields highest MRR).}
    \begin{tabular}{|c|c|c|c|c|}
     \hline
    \label{tbl:experiment_results}
    \textbf{Architecture} &  \textbf{Augmented Scenario} & \textbf{Source} & \textbf{MRR} \\
    \hline  
    CSNet & NBOW, ACS=\#0  & [1] & 0.580 \\
    \hline 
    CSNet & CNN, ACS=\#0  & [1] & 0.573 \\
    \hline 
    CSNet & BIRNN, ACS=\#0  & [1] & 0.321 \\
    \hline 
    CSNet & SelfAtt, ACS=\#0  & [1] & 0.692 \\
    \hline 
    CSNet & SelfAtt, ACS=\#1 & Our & 0.693 \\
    \hline 
    CSNet & SelfAtt , ACS=\#2 & Our & 0.710  \\
    \hline 
    CSNet & SelfAtt, ACS=\#3 &  Our  & 0.727 \\
    \hline
    CSNet & SelfAtt , \textbf{ACS=\#4} & Our & \textbf{0.734} \\ 
\hline 
CSNet & SelfAtt, ACS=\#5 & Our & 0.670 \\
\hline \hline
CodeBERT & RoBERTa, ACS=\#0  & [2] & 0.808 \\
\hline 
CodeBERT & PT W/Code Only (INIT=S), ACS=\#0  & [2] & 0.785 \\
\hline
CodeBERT & PT W/Code Only (INIT=R), ACS=\#0  & [2] & 0.843  \\
\hline 
CodeBERT & MLM, INIT=S,ACS=\#0  & [2] & 0.826 \\
\hline 
CodeBERT & MLM, INIT=R, ACS=\#0  & [2] & 0.864 \\
\hline 
CodeBERT & RTD, INIT=R, ACS=\#0   & [2] & 0.826 \\
\hline 
CodeBERT & MLM+RTD, INIT=R, ACS=\#0   & [2] & 0.868  \\
\hline 
CodeBERT & MLM+RTD, INIT=R, \textbf{ACS=\#4}  & Our & \textbf{0.961} \\
\hline 
    \end{tabular}
\end{table}
As we explained previously, the scalability of code retrieval model is important. Mainly previous studies only focused on small portion of testing dataset (999 distractor code). In order to examine the scalability of the best outperformed model, we considered different magnitude of datasets as search space (batch size).  
Table~\ref{tbl:experiment_results_sizes} shows the experimental results on different sizes by using the most outperformed model from Table~\ref{tbl:experiment_results} (last row). In this experiment, \textit{search space} refers to the number of records in target search dataset, \textit{Magnitude} shows the search size expansion against original code search which is used in CodeSearchNet and CodeBERT architectures, \textit{Number of Queries} refers to the total number of given queries in each experiment, \textit{Top(1) MRR}. 
The model has been trained on augmented programming language code rather than simply considering the abstract or a summary of the docstring.
This results shows that even with increasing the target search dataset (e.g., 45x greater than the original 999 distractor code search space size, which includes both testing and validation datasets), the model was able to retrieve correct source-code with a reasonable performance.  Since each experiment selected a set of query from its dataset (shuffled randomly for fairness), then the performance might be relied on the given query and its rank in MRR computation. 
\begin{table}[h]
    \centering
    \caption{Experiment results on different search space  (ACS=4)}
    \begin{tabular}{|c|c|c|c|c|c|}
        \hline
        \label{tbl:experiment_results_sizes}
        \textbf{Search Space} &
        \textbf{Magnitude} &
        \textbf{Number of}  &
        \textbf{Top (1)}  \\ 
         &  & \textbf{ Queries} &\textbf{MRR} \\
        \hline  
        22,176 & 22x & 45 & 0.840  \\
        \hline  
        23,107 & 23x & 45 & 0.823  \\
        \hline  
        45,283 & \textbf{45x} & 21 & 0.821 \\
        \hline  
    \end{tabular}
\end{table}
Another evaluation approach is micro matching evaluation which is defined as follows. After applying only positive examples from testing dataset it aims to predict correct/incorrect matching between docstring and code from testing dataset.
We used the best outperformed model as described in Table~\ref{tbl:experiment_results}, and run infer on all test dataset that includes 22,176 sample data from testing dataset. We consider accuracy as $Acc\footnotesize{\%}=\frac{T_P}{T_Q}$ where $T_P$ represents the total number of True-Positive matched of code to docstring, and $T_Q$ represents the total number of queries. Submitted queries to the model for each pair of function and description returns an integer between $[-4.33,5.19]$ (per our experiment setup) where the output of greater than zero considered as matched record. The output shows that the trained model based on augmented programming language model of \textit{Scenario\#4} was able to predict correctly all positive matched records with an accuracy of \textbf{99.21\%}.
\section{Demonstration}
\label{sec:demonstration}
We fine-tuned CodeBERT based on the best augmented code \textit{$ACS$=4} and generated an API on top of model that showcases the outperformed model over other augmented code scenarios with an MRR of 0.96. The model is published in HuggingFace Hub as \href{https://huggingface.co/Fujitsu/AugCode/tree/main}{AugCode} and a demonstration video is published in \href{https://youtu.be/mnZrUTANjGs}{YouTube}. The framework is designed as a Command Line Interface (CLI) in Ubuntu which i) takes a natural language query input (e.g., "\textit{save a csv file}"), ii) select a model and the search query space where it allows the user to generate a search space (batch) of 1x, or 45x that corresponds to 999 distractor codes from testing dataset,  testing dataset + validation dataset, respectively; iii) perform inferences on search space to find best candidate code for given natural language query, iv) shows top 10 retrieved codes as output. The API can be easily plug-in to an IDE (e.g., Jupyter Notebook) where a user may type a query and get top 1 or top 10 retrieved codes as suggestion for each user query, that may help user to quickly and efficiently implement his/her code.    
\section{Conclusion}
\label{sec:conclusion}
This paper analysed different attributes of code retrieval by introducing \textit{AugmentedCode}. We utilized two recent machine-learning based architectures (CodeSearchNet and CodeBERT) to compare different augmented code scenarios on a large number of curated data records. Our fine-tuned model achieved an MRR of 0.961. Our results show that a full code description (entire docstring) including parameters definitions, interestingly with codes' comments significantly improved the performance of the state-of-the-art code retrieval models. Adding additional information such as GitHub commit messages also improved the baseline performance.
\bibliographystyle{plain}
\bibliography{refs}

\begin{thebibliography}{10}

\bibitem{pytorrent-v1-model}
Pytorrent transformer-based language model.
\newblock \url{https://huggingface.co/Fujitsu/pytorrent}.

\bibitem{pytorrent-zenodo}
Pytorrent datasets.
\newblock \url{https://doi.org/10.5281/zenodo.4451357}, 2021.
\newblock Accessed: 10/16/2021.

\bibitem{pytorrent_github}
Pytorrent github page.
\newblock \url{https://github.com/fla-sil/PyTorrent}, 2021.
\newblock Accessed: 10/16/2021.

\bibitem{bahrami2021pytorrent}
Mehdi Bahrami, N~C Shrikanth, Shade Ruangwan, , Lei Liu, Yuji Mizobuchi,
  Masahiro Fukuyori, Wei-Peng Chen, Tim Menzies, and Kazuki Munakata.
\newblock Pytorrent: A python library corpus for large-scale language models.
\newblock {\em \url{https://arxiv.org/pdf/2110.01710}}, 2021.

\bibitem{cambronero2019deep}
Jose Cambronero, Hongyu Li, Seohyun Kim, Koushik Sen, and Satish Chandra.
\newblock When deep learning met code search.
\newblock In {\em Proceedings of the 2019 27th ACM Joint Meeting on European
  Software Engineering Conference and Symposium on the Foundations of Software
  Engineering}, pages 964--974, 2019.

\bibitem{devlin2018bert}
Jacob Devlin, Ming-Wei Chang, Kenton Lee, and Kristina Toutanova.
\newblock Bert: Pre-training of deep bidirectional transformers for language
  understanding.
\newblock {\em arXiv preprint arXiv:1810.04805}, 2018.

\bibitem{endres2003handbook}
Albert Endres and H~Dieter Rombach.
\newblock {\em A handbook of software and systems engineering: Empirical
  observations, laws, and theories}.
\newblock Pearson Education, 2003.

\bibitem{feng2020codebert}
Zhangyin Feng, Daya Guo, Duyu Tang, Nan Duan, Xiaocheng Feng, Ming Gong, Linjun
  Shou, Bing Qin, Ting Liu, Daxin Jiang, et~al.
\newblock Codebert: A pre-trained model for programming and natural languages.
\newblock {\em arXiv preprint arXiv:2002.08155}, 2020.

\bibitem{hindle2012naturalness}
Abram Hindle, Earl~T Barr, Zhendong Su, Mark Gabel, and Premkumar Devanbu.
\newblock On the naturalness of software.
\newblock In {\em 2012 34th International Conference on Software Engineering
  (ICSE)}, pages 837--847. IEEE, 2012.

\bibitem{hu2018deep}
Xing Hu, Ge~Li, Xin Xia, David Lo, and Zhi Jin.
\newblock Deep code comment generation.
\newblock In {\em 2018 IEEE/ACM 26th International Conference on Program
  Comprehension (ICPC)}, pages 200--20010. IEEE, 2018.

\bibitem{husain2019codesearchnet}
Hamel Husain, Ho-Hsiang Wu, Tiferet Gazit, Miltiadis Allamanis, and Marc
  Brockschmidt.
\newblock Codesearchnet challenge: Evaluating the state of semantic code
  search.
\newblock {\em arXiv preprint arXiv:1909.09436}, 2019.

\bibitem{lebeuf2017software}
Carlene Lebeuf, Margaret-Anne Storey, and Alexey Zagalsky.
\newblock Software bots.
\newblock {\em IEEE Software}, 35(1):18--23, 2017.

\bibitem{liu2020opportunities}
Chao Liu, Xin Xia, David Lo, Cuiyun Gao, Xiaohu Yang, and John Grundy.
\newblock Opportunities and challenges in code search tools.
\newblock {\em arXiv preprint arXiv:2011.02297}, 2020.

\bibitem{liu2019roberta}
Yinhan Liu, Myle Ott, Naman Goyal, Jingfei Du, Mandar Joshi, Danqi Chen, Omer
  Levy, Mike Lewis, Luke Zettlemoyer, and Veselin Stoyanov.
\newblock Roberta: A robustly optimized bert pretraining approach.
\newblock {\em arXiv preprint arXiv:1907.11692}, 2019.

\bibitem{peters2018deep}
Matthew~E Peters, Mark Neumann, Mohit Iyyer, Matt Gardner, Christopher Clark,
  Kenton Lee, and Luke Zettlemoyer.
\newblock Deep contextualized word representations.
\newblock {\em arXiv preprint arXiv:1802.05365}, 2018.

\bibitem{radford2018improving}
Alec Radford, Karthik Narasimhan, Tim Salimans, and Ilya Sutskever.
\newblock Improving language understanding with unsupervised learning.
\newblock {\em Technical report, OpenAI}, 2018.

\bibitem{sadowski2015developers}
Caitlin Sadowski, Kathryn~T Stolee, and Sebastian Elbaum.
\newblock How developers search for code: a case study.
\newblock In {\em Proceedings of the 2015 10th Joint Meeting on Foundations of
  Software Engineering}, pages 191--201, 2015.

\bibitem{shrikanth2020assessing}
NC~Shrikanth and Tim Menzies.
\newblock Assessing practitioner beliefs about software defect prediction.
\newblock In {\em Proceedings of the ACM/IEEE 42nd International Conference on
  Software Engineering: Software Engineering in Practice}, pages 182--190,
  2020.

\bibitem{singer2010examination}
Janice Singer, Timothy Lethbridge, Norman Vinson, and Nicolas Anquetil.
\newblock An examination of software engineering work practices.
\newblock In {\em CASCON First Decade High Impact Papers}, pages 174--188.
  2010.

\bibitem{sridhara2011automatically}
Giriprasad Sridhara, Lori Pollock, and K~Vijay-Shanker.
\newblock Automatically detecting and describing high level actions within
  methods.
\newblock In {\em 2011 33rd International Conference on Software Engineering
  (ICSE)}, pages 101--110. IEEE, 2011.

\end{thebibliography}
\end{document}